\documentclass[conference]{IEEEtran}
\usepackage{amsmath}
\usepackage{array}
\usepackage{graphicx}       
\usepackage{subcaption}     
\IEEEoverridecommandlockouts

\usepackage{cite}
\usepackage{amsmath,amssymb,amsfonts}
\usepackage{algorithmic}
\usepackage{graphicx}
\usepackage{textcomp}
\usepackage{xcolor}
\def\BibTeX{{\rm B\kern-.05em{\sc i\kern-.025em b}\kern-.08em
    T\kern-.1667em\lower.7ex\hbox{E}\kern-.125emX}}

\begin{document}

\title{Multi-Objective Routing Optimization Using Coherent Ising Machine in Wireless Multihop Networks}

\author{
	\IEEEauthorblockN{Yuxuan Lin, Chuyao Xu, and Chuan Wang\textsuperscript{\textdagger}}
	\IEEEauthorblockA{
		\textit{School of Artificial Intelligence}, \textit{Beijing Normal University}, Beijing, China \\
		\{linyx, xuchuyao\}@mail.bnu.edu.cn, wangchuan@bnu.edu.cn
	}
	\thanks{\textsuperscript{\textdagger}Corresponding author: Chuan Wang (wangchuan@bnu.edu.cn).}
}

\maketitle

\begin{abstract}

Multi-objective combinatorial optimization in wireless communication networks is a challenging task, particularly for large-scale and diverse topologies. Recent advances in quantum computing offer promising solutions for such problems. Coherent Ising Machines (CIM), a quantum-inspired algorithm, leverages quantum properties of coherent light, enabling faster convergence to the ground state. This paper applies CIM to multi-objective routing optimization in wireless multi-hop networks. We formulate the routing problem as a Quadratic Unconstrained Binary Optimization (QUBO) problem, and map it onto an Ising model, allowing CIM to solve it. CIM demonstrates strong scalability across diverse network topologies without requiring topology-specific adjustments, overcoming the limitations of traditional quantum algorithms like Quantum Approximate Optimization Algorithm (QAOA) and Variational Quantum Eigensolver (VQE). Our results show that CIM provides feasible and near-optimal solutions for networks containing hundreds of nodes and thousands of edges. Additionally, a complexity analysis highlights CIM’s increasing efficiency as network size grows.
	
\end{abstract}


\section{Introduction}
Wireless communication networks have experienced rapid advancements, driving significant technological progress across various sectors. As networks grow in size and complexity, the demand for faster data transmission and more efficient routing becomes increasingly important. Traditional computational methods are effective for single-objective optimization but struggle with multi-objective combinatorial optimization, particularly in large-scale networks. These problems involve the balancing of competing objectives, making them NP-hard and challenging to solve.

Quantum computing has already demonstrated potential in addressing some of these challenges in wireless communication\cite{bib1}, particularly in areas such as resource allocation\cite{bib2}, signal processing\cite{bib3}, and network optimization\cite{bib4}. By leveraging quantum superposition and entanglement, quantum algorithms can explore multiple solutions simultaneously, offering potential advantages in solving complex combinatorial problems\cite{bib5}\cite{bib6}. Algorithms such as the Quantum Approximate Optimization Algorithm (QAOA) and the Variational Quantum Eigensolver (VQE) have been applied to small-scale network routing problems\cite{bib7}\cite{bib8}\cite{bib9}\cite{bib10}, demonstrating the potential for improving efficiency. However, despite their advantages, these algorithms face scalability issues when applied to larger networks with complex topologies. As the problem size increases, the quantum circuits required by QAOA and VQE become more complex, resulting in significant hardware constraints and longer execution times. Additionally, these algorithms often require problem-specific circuit designs and extensive parameter tuning, limiting their adaptability across diverse network scenarios.

To address these challenges, we propose the application of Coherent Ising Machines (CIM)\cite{bib11}, a quantum-inspired computational approach, for solving multi-objective routing optimization in wireless multi-hop networks. CIM leverages quantum properties of coherent light to efficiently solve the problem via Ising model\cite{bib12}. Unlike QAOA and VQE, CIM offers greater scalability and flexibility, as it does not require topology-specific circuit adjustments, making it well-suited for large-scale and diverse network environments.

The key contributions of this work are as follows:
\begin{itemize}
	\item We introduce the first application of CIM to multi-objective optimization in wireless multi-hop networks, demonstrating its capability to solve routing problems across various network topologies. This contrasts sharply with other quantum methods, which often need custom adjustments for each topology.
	\item Our results demonstrate that CIM is capable of finding feasible, and even optimal, solutions for networks with hundreds of nodes and thousands of edges. A detailed complexity analysis further highlights the superiority of CIM in terms of computational efficiency as the network size increases.
\end{itemize}

The remainder of this paper is structured as follows: Section 2 introduces our network routing model and its transformation into an Ising model via the QUBO model. Section 3 outlines the principles of CIM and the mapping process from the Ising model to CIM. Section 4 presents a comprehensive analysis of the experimental results, including comparisons of scalability and computational complexity. Finally, Section 5 concludes the paper and discusses future research directions.

\begin{figure}[htbp]
	\centering
	\includegraphics[scale=0.65]{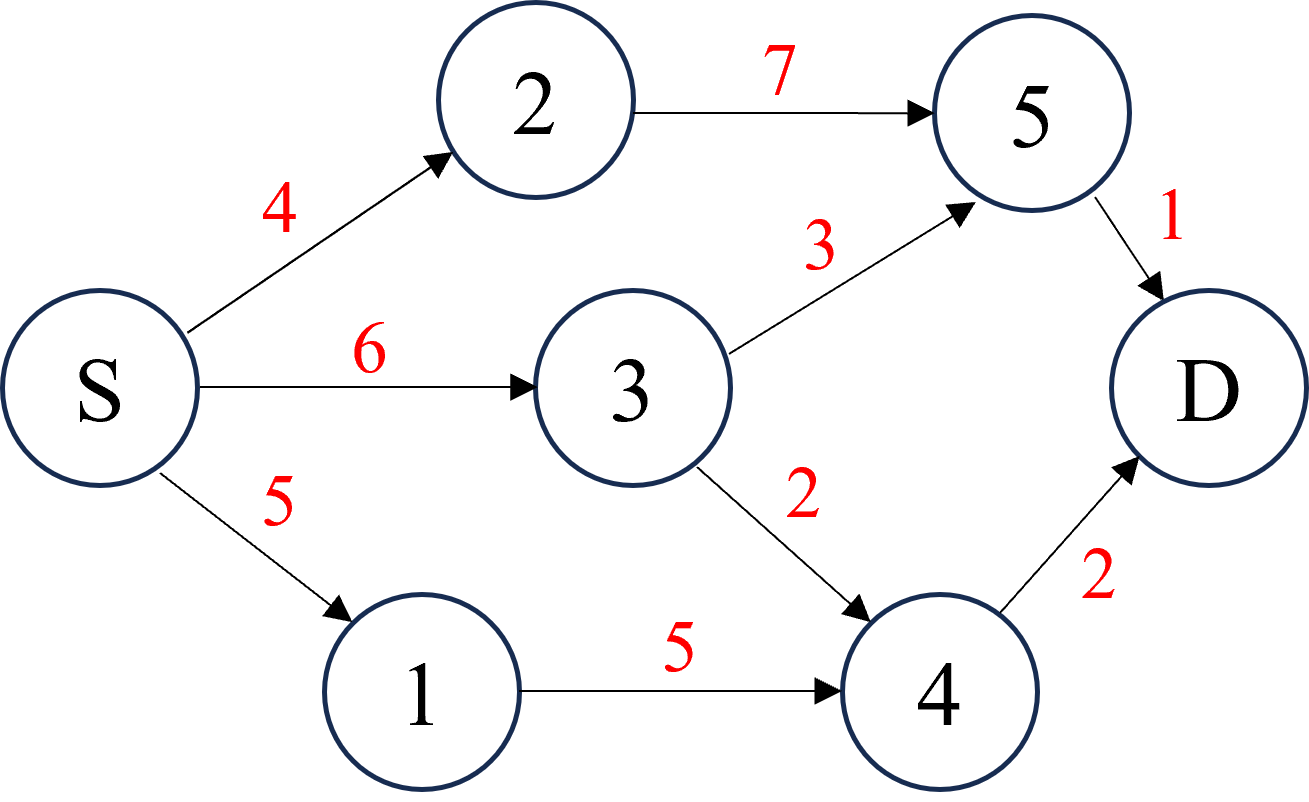}
	\caption{Example of a weighted directed graph.}
	\label{fig1}
\end{figure}

\section{System Model and Problem Formulation}
In this section, we present our network model and formulate it mathematically, then transform the formulation into the Ising model via the QUBO model.
\subsection{Network model}\label{AA}
We consider a wireless multi-hop network modeled as a directed graph $G=(V,E,W)$\cite{bib13}, which comprises a set of nodes $V$, directed edges $E$, and edge weights $w_{i,j}$ representing the cost of transmitting data across the edge from node $i$ to node $j$. The source node is denoted by $S$, while the destination node is denoted by $D$. Fig. \ref{fig1} illustrates an example of such a weighted directed graph.

We define a binary variable $x_{i,j}$, where $x_{i,j} = 1$ if the directed edge from node $i$ to node $j$ is included in the path, and $x_{i,j} = 0$ otherwise. The basic routing problem can be expressed as follows:
\begin{equation}
	\underset{x}{\text{argmin}} \, f(x)
	\label{label_1}
\end{equation}
\begin{equation}
	\sum_j x_{i,j} - \sum_k x_{k,i} = 
	\begin{cases}
		1, & \text{if } i = S, \\
		-1, & \text{if } i = D, \\
		0, & \text{otherwise}.
	\end{cases}
	\label{label_2}
\end{equation}

These constraints don't define a strict shortest-path problem. To simplify the problem, we restrict the in-degree of the source node and the out-degree of the destination node, transforming (\ref{label_2}) into:
\begin{equation}
\label{label_3}
\begin{cases}
	\sum_jx_{i,j}=1,&\quad\mathrm{if~}i=S\\\sum_kx_{k,i}=1,&\quad\mathrm{if~}i=D\\\sum_jx_{i,j}-\sum_kx_{k,i}=0,&\quad\mathrm{otherwise}
\end{cases}
\end{equation}

We aim to optimize multiple objectives: path loss, bit error rate(BER), and hop count\cite{bib14}\cite{bib15}.
\paragraph{Path Loss}\mbox{}
Path loss quantifies the attenuation of electromagnetic signals between nodes, increasing with distance. In this work, it is modeled deterministically, omitting the stochastic variations typical of wireless environments\cite{bib14}. This assumption facilitates a more tractable mathematical analysis and improves computational efficiency, particularly for large-scale optimization tasks. It is given by:
\begin{equation}
	L_{ij}=\frac{P_{T}}{P_{R,ij}}=\left(\frac{4\pi d_{ij}}{\lambda_c}\right)^\alpha \label{eq}
\end{equation}
where $P_T$ is the transmitted power, $P_{R,ij}$ is the received power at node $j$, $\lambda_c$ is the carrier wavelength, and $d_{ij}$ is the Euclidean distance between nodes $i$ and $j$. The total path loss can be expressed as a sum of linear terms:

\begin{equation}
	E_{\mathrm{LOSS}}(x)=\sum\limits_{(i,j)\in E}L_{ij}x_{ij}.
\end{equation}

\paragraph{Bit Error Rate}\mbox{}
The BER indicates the probability of error during transmission, influenced by signal power and noise\cite{bib14}. It is given by:
\begin{equation}
	p_{ij} = \frac{1}{2} \left( 1 - \sqrt{\frac{R_{ij}}{R_{ij} + 1}} \right)
\end{equation}
where $R_{ij}$ is the signal-to-noise ratio (SNR):
\begin{equation}
	R_{ij} = \frac{P_{R,ij}}{\log_2(M) P_{N,ij}}
\end{equation}
with \(M\) representing the modulation scheme (e.g., \(M = 2\) for BPSK, \(M = 4\) for QPSK), and $P_{N,ij}$ is the average noise power at nodes $i$ and $j$, defined as:
\begin{equation}
	P_{N,ij} = \frac{1}{2} (P_{N,i} + P_{N,j}).
\end{equation}
here, we assume the noise $P_{N,i}$ follows a normal distribution with mean $\mu = -90$ dBm and standard deviation $\sigma = 10$ dBm. The total BER is the sum of the individual error probabilities for each selected edge. It is a first-order approximation of the exact total error (equivalent to assuming errors do not occur twice on the same bit). The total BER is:

\begin{equation}
	E_{\mathrm{BER}}(x) = \sum_{(i,j) \in E} p_{ij} x_{ij}.
\end{equation}

\paragraph{Hops}\mbox{}
The number of hops represents the total number of edges used in the path and is given by:
\begin{equation}
	E_{\mathrm{HOP}}(x) = \sum_{(i,j) \in E} h_{ij} x_{ij}
\end{equation}
where $h_{ij}$ represents the weight associated with each edge, typically considered equal for counting hops.

We aim to minimize path loss, bit error, and the number of hops simultaneously. These objectives often conflict, so it is unlikely to find a single solution that optimizes all objectives. Instead, we seek a set of Pareto optimal solutions\cite{bib16}, where no solution can improve one objective without worsening another. We use a scalarization technique\cite{bib17} to aggregate these objectives into a single function:
\begin{equation}
	f(x) = v_1 E_{\mathrm{LOSS}}(x) + v_2 E_{\mathrm{BER}}(x) + v_3 E_{\mathrm{HOP}}(x)
\end{equation}
where $v_1$, $v_2$, and $v_3$ are weighting coefficients that satisfy $\sum_{i=1}^{3} v_i = 1$. The optimization problem thus becomes finding the minimum of $f(x)$, subject to the constraints in (\ref{label_3}).

\subsection{Quadratic Unconstrained Binary Optimization Model}
The Quadratic Unconstrained Binary Optimization (QUBO) model provides a framework for solving combinatorial optimization problems, where the objective is to determine the optimal configuration of binary variables that minimizes a quadratic objective function. Formulated as an unconstrained optimization problem, it encodes the constraints within the objective function. The model is defined over binary variables \(x_i \in \{0, 1\}\), and uses the quadratic function \(H(x)\) to represent pairwise interactions:

\begin{equation}
	H(x) = \sum_{i \neq j} Q_{ij} x_i x_j + \sum_i q_i x_i
\end{equation}
where \(Q_{ij}\) denotes the interaction coefficients between \(x_i\) and \(x_j\), and \(q_i\) represents the linear coefficients for \(x_i\). The objective function \(H(x)\) can be interpreted as a Hamiltonian, analogous to the energy function in Ising model.

To incorporate constraints from (\ref{label_3}), we add penalty terms:
\begin{align}
	\label{label_13}
	H(x) &= f(x) + P_1 \left( \sum_i x_{S,i} - 1 \right)^2 
	+ P_2 \left( \sum_i x_{i,D} - 1 \right)^2 \notag \\
	&\quad + P_3 \sum_{i \not\in \{S,D\}} \left( \sum_j x_{i,j} - \sum_k x_{k,i} \right)^2 
\end{align}
where \(P_1\), \(P_2\), and \(P_3\) are positive penalty coefficients that must be set large enough to ensure that any violation of the constraints is penalized heavily. Typically, \(P_3\) is chosen to be larger than \(P_1\) and \(P_2\) to enforce constraints more strictly.

Expanding (\ref{label_13}) yields the quadratic terms \(Q_{ij}\) and linear terms \(q_i\) as follows:
\begin{equation}
	\begin{aligned}
		Q_{ij} = & \, P_1 \sum_{i \neq j} x_{S,i} x_{S,j} + P_2 \sum_{i \neq j} x_{i,D} x_{j,D} \\
		& + P_3 \sum_{i \not\in \{S,D\}} \left( \sum_{(i,j) \in E} \sum_{\substack{(i,k) \in E \\ k \neq j}} x_{i,j} x_{i,k} \right. \\
		& \quad + \left. \sum_{(j,i) \in E} \sum_{\substack{(k,i) \in E \\ k \neq j}} x_{j,i} x_{k,i} 
		- 2 \sum_{(i,j) \in E} \sum_{(k,i) \in E} x_{i,j} x_{k,i} \right)
	\end{aligned}
\end{equation}
\begin{equation}
	\begin{aligned}
		q_i = & -P_1 \sum_i x_{S,i} - P_2 \sum_i x_{i,D} \\
		& + P_3 \sum_{i \not\in \{S,D\}} \left( \sum_{(i,j) \in E} x_{i,j} + \sum_{(k,i) \in E} x_{k,i} \right) + f(x)
	\end{aligned}
\end{equation}

\subsection{Ising Model}
The Ising model is a mathematical representation in statistical physics used to describe magnetic materials, where the system's energy depends on the configuration of spin variables. In the Ising model, each spin variable \(\sigma_i \in \{-1,1\}\) represents a binary state. The model's Hamiltonian is given by:
\begin{equation}
	H(\sigma) = -\sum_{i,j} J_{ij} \sigma_i \sigma_j - \sum_i h_i \sigma_i
\end{equation}
where \(J_{ij}\) denotes the coupling strength between spins \(i\) and \(j\), and \(h_i\) represents the influence of an external magnetic field on spin \(i\).

To convert the QUBO model to the Ising model, we use the transformation \(\sigma_i = 2x_i - 1\), which maps the binary variables \(x_i \in \{0,1\}\) to Ising spins \(\sigma_i \in \{-1,1\}\). The QUBO expression can be rewritten in terms of Ising variables:

\begin{equation}
	\begin{aligned}
		H(\sigma) &= \sum_{i \neq j} Q_{ij} \cdot \frac{(\sigma_i + 1)(\sigma_j + 1)}{4} + \sum_i q_i \cdot \frac{\sigma_i + 1}{2} \\
		&= \sum_{i \neq j} \frac{Q_{ij}}{4} \sigma_i \sigma_j + \sum_i \left( \frac{q_i}{2} + \frac{\sum_{i \neq j} (Q_{ij} + Q_{ji})}{4} \right) \sigma_i \\
		&\quad + \left( \frac{\sum_i q_i}{2} + \frac{\sum_{i \neq j} Q_{ij}}{4} \right)
	\end{aligned}
\end{equation}
Thus, the coefficients in the Ising model are:
\begin{equation}
	J_{ij} = -\frac{Q_{ij}}{4}
\end{equation}
\begin{equation}
	h_i = -\frac{q_i}{2} - \frac{\sum_{i \neq j} (Q_{ij} + Q_{ji})}{4}
\end{equation}

We use this mapping to transform the original optimization problem into the Ising model, allowing it to be solved using CIM.

\section{Coherent Ising Machine}

In this section, we introduce Coherent Ising Machine (CIM) and explain its application to solving our multi-objective routing model.

CIM is a quantum-inspired optimization technique that leverages principles from the Ising model. The Ising model represents optimization problems as energy minimization tasks, where the goal is to find the ground state of a system by minimizing its energy. In the context of combinatorial optimization, this ground state corresponds to an optimal or near-optimal solution. CIM utilizes an optical network of degenerate optical parametric oscillators (DOPOs) to represent spins in the Ising model. Each DOPO encodes the spin states using phase-coherent laser pulses, with the two possible phases corresponding to spin values of $+1$ or $-1$. The system's energy is minimized by continuously adjusting the pump power, allowing CIM to evolve the spin states toward lower energy configurations and find solutions to the optimization problem.

Theoretically, CIM operates by solving an Ising problem defined by a Hamiltonian that represents the energy of the spin system. The objective is to find a spin configuration that minimizes this Hamiltonian, which corresponds to solving a combinatorial optimization problem. In practical implementations, the dynamic behavior of CIM can be modeled using differential equations that describe the time evolution of the in-phase and quadrature-phase components of the DOPO amplitudes. While traditional studies often neglect the noise term in these equations\cite{bib12}, we incorporate the first-order term matrix into the noise term to better capture the stochastic nature of the system\cite{bib2}. It allows CIM to efficiently explore the solution space of NP-hard problems, making it a powerful tool for multi-objective routing optimization.

\subsection{Dynamic Equations of CIM}
The dynamic behavior of CIM can be described using stochastic differential equations that model the time evolution of the in-phase ($c_i$) and quadrature-phase ($s_i$) components of the amplitude of each DOPO. The coupled differential equations for these components are given by:

\begin{equation}
\label{label_20}
	\frac d{dt}c_i=(-1+p-(c_i^2+s_i^2))c_i+\sum_{j=1,j\neq i}^NJ_{ij}c_j-h_{i}
\end{equation}
\begin{equation}
	\label{label_21}
	\frac d{dt}s_i=(-1-p-(c_i^2+s_i^2))s_i+\sum_{j=1,j\neq i}^NJ_{ij}s_j-h_{i}
\end{equation}

In these equations:
\begin{itemize}
	\item $c_i$ and $s_i$ denote the in-phase and quadrature-phase components of the $i$-th DOPO. The Ising spin $\sigma_i$ is determined by $c_i$. If $c_i<0$, then  $\sigma_i= -1$ and if $c_i \ge 0$, then $\sigma_i = +1 $. 
	\item $J_{ij}$ represents the coupling coefficient between spins $i$ and $j$, encoding the strength of interactions between the corresponding variables.
	\item The hyper-parameter $p$ serves as the pump pulse in the system, used to amplify the in-phase component $c_i$. To ensure efficient convergence to the ground state, the pump parameter $p(t)$ is dynamically adjusted over time, which is defined as:
	\begin{equation}
		p(t) = p(t-1) \cdot \tanh(0.0005 \cdot t)
	\end{equation}
	This recursive update allows the system to gradually increase the pump power, which helps CIM explore the solution space more effectively, thereby increasing the probability of finding the global optimum.
\end{itemize}

\begin{figure*}[htbp]
	\centering
	\begin{subfigure}[b]{0.32\textwidth}
		\centering
		\includegraphics[width=\textwidth]{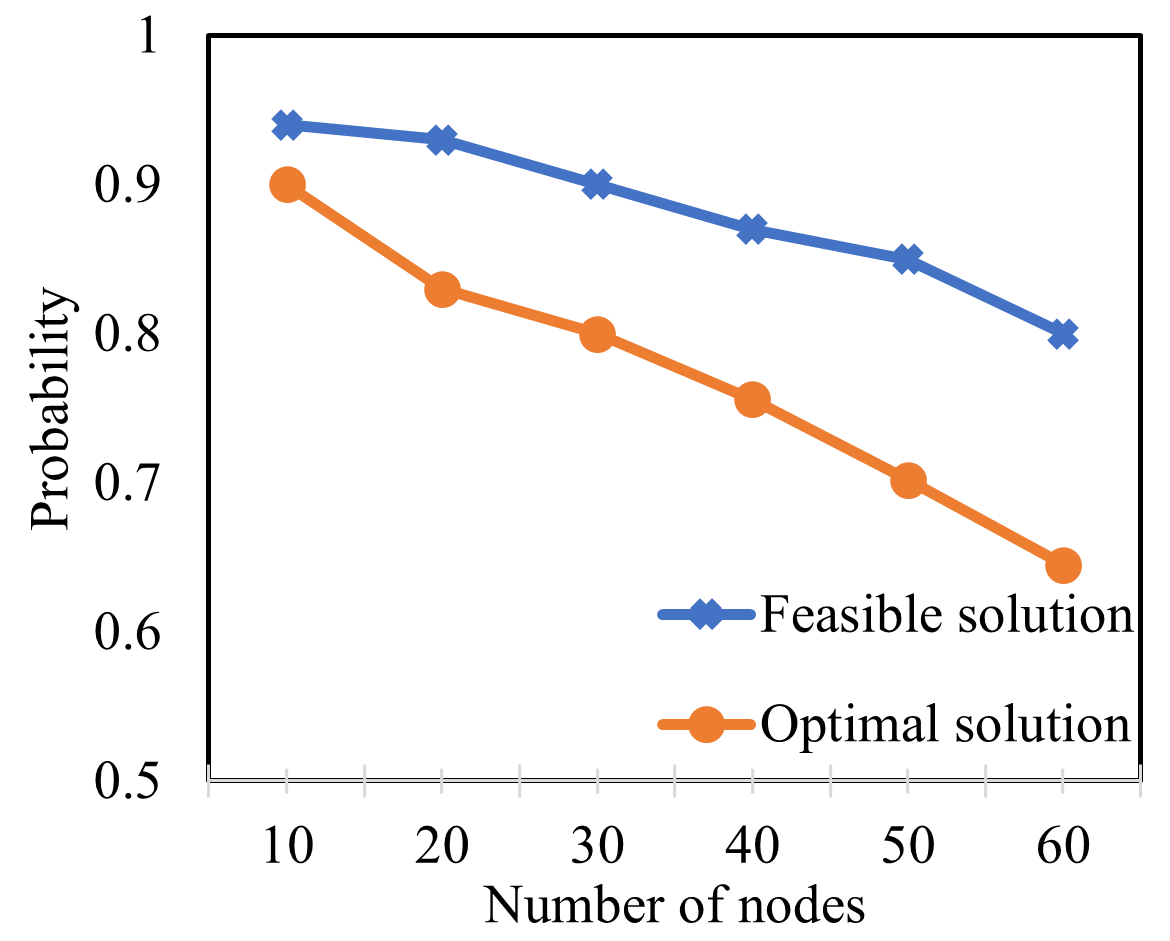}
		\caption{\hspace{0.25cm}Path loss}
		\label{fig2.1}
	\end{subfigure}
	\hfill
	\begin{subfigure}[b]{0.32\textwidth}
		\centering
		\includegraphics[width=\textwidth]{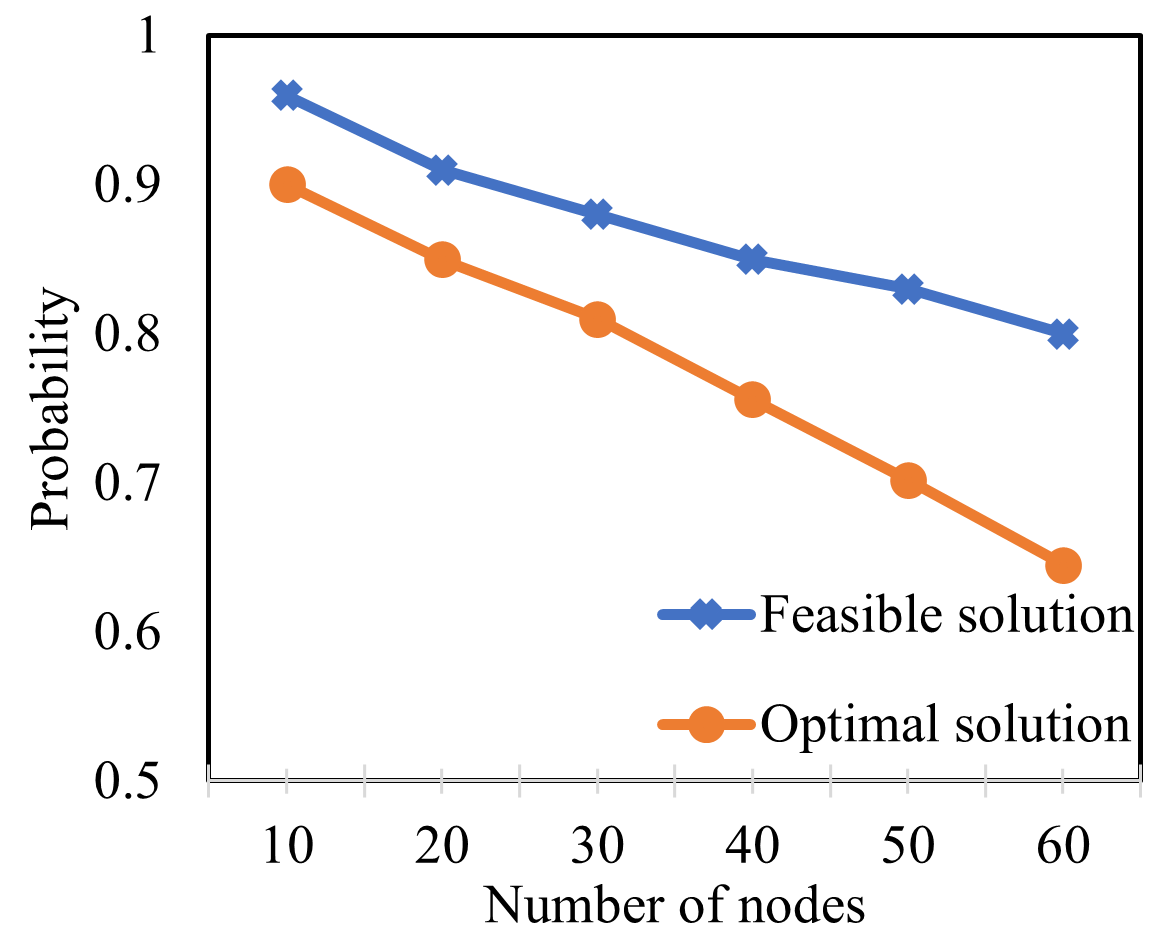}
		\caption{\hspace{0.25cm}BER}
		\label{fig2.2}
	\end{subfigure}
	\hfill
	\begin{subfigure}[b]{0.32\textwidth}
		\centering
		\includegraphics[width=\textwidth]{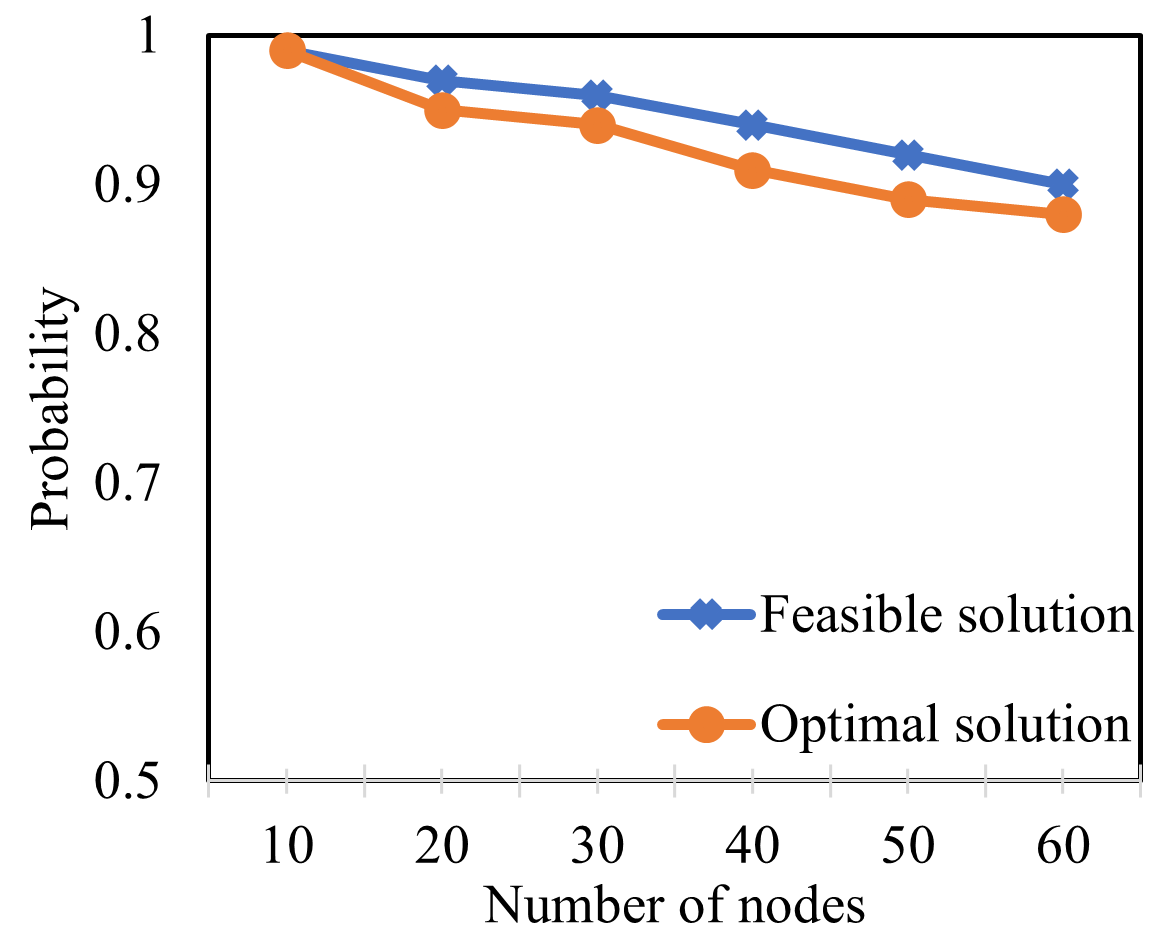}
		\caption{\hspace{0.25cm}Hops}
		\label{fig2.3}
	\end{subfigure}
	
	\caption{Probability of feasible solutions and optimal solutions for a single objective}
	\label{fig2}
\end{figure*}

\begin{figure*}[htbp]
	\centering
	\begin{subfigure}[b]{0.32\textwidth}
		\centering
		\includegraphics[width=\textwidth]{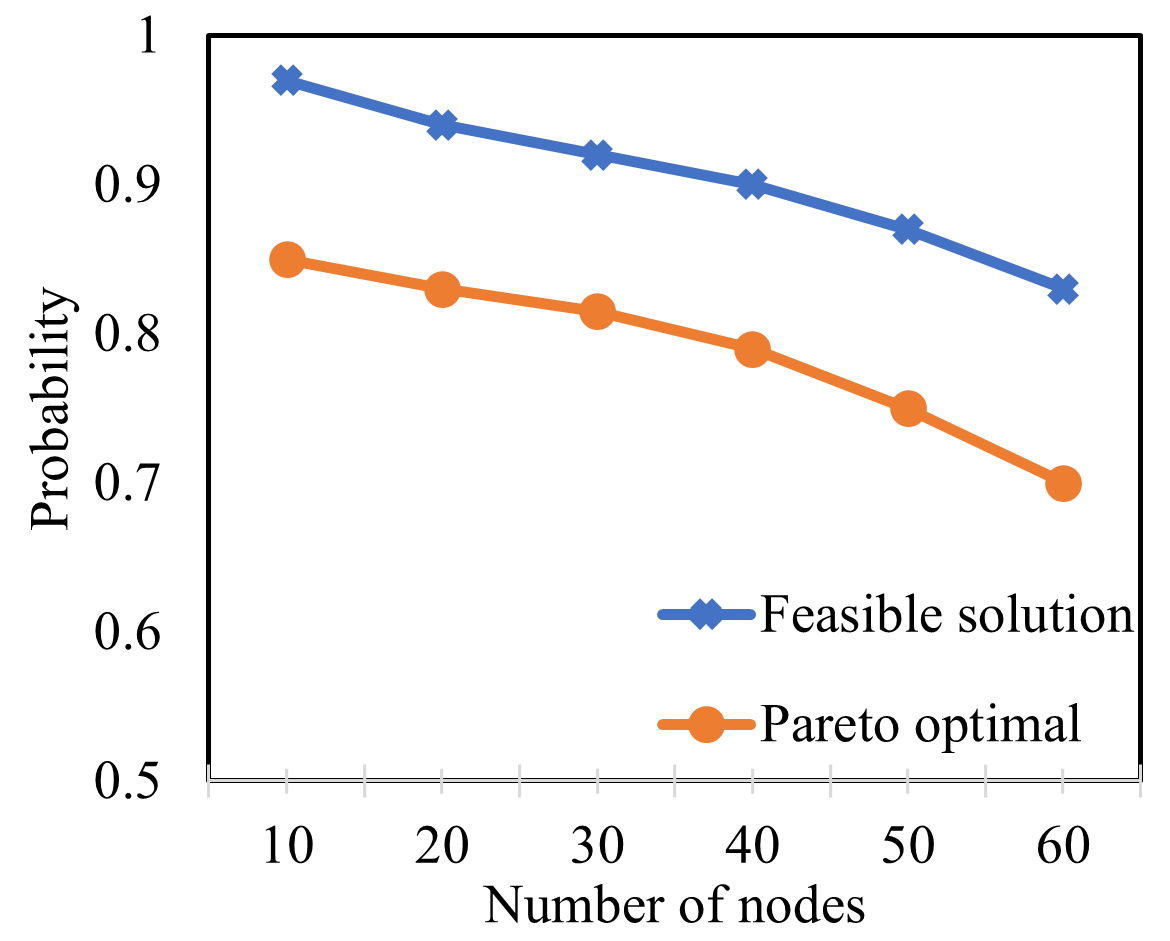}
		\caption{\hspace{0.25cm}Path loss and BER}
		\label{fig3.1}
	\end{subfigure}
	\hfill
	\begin{subfigure}[b]{0.32\textwidth}
		\centering
		\includegraphics[width=\textwidth]{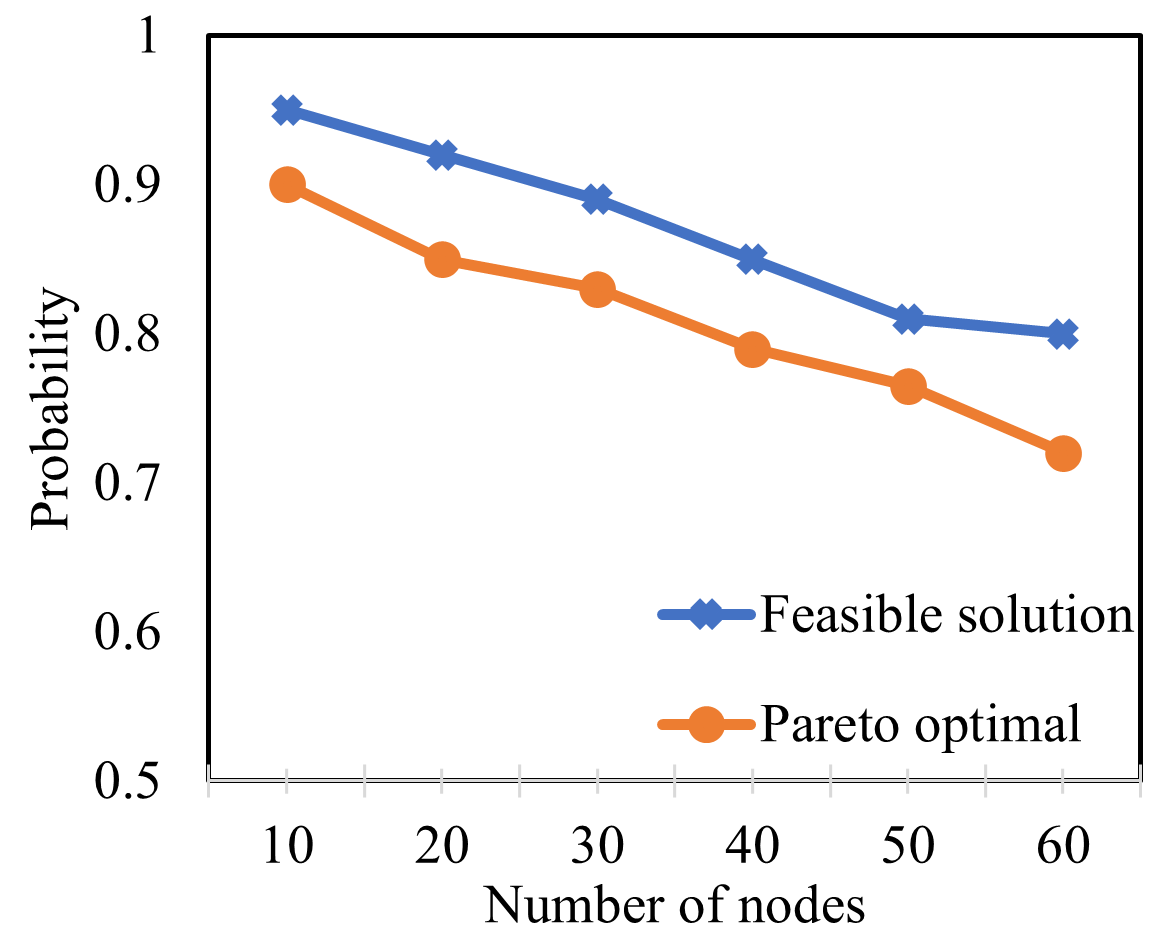}
		\caption{\hspace{0.25cm}Path loss and hops}
		\label{fig3.2}
	\end{subfigure}
	\hfill
	\begin{subfigure}[b]{0.32\textwidth}
		\centering
		\includegraphics[width=\textwidth]{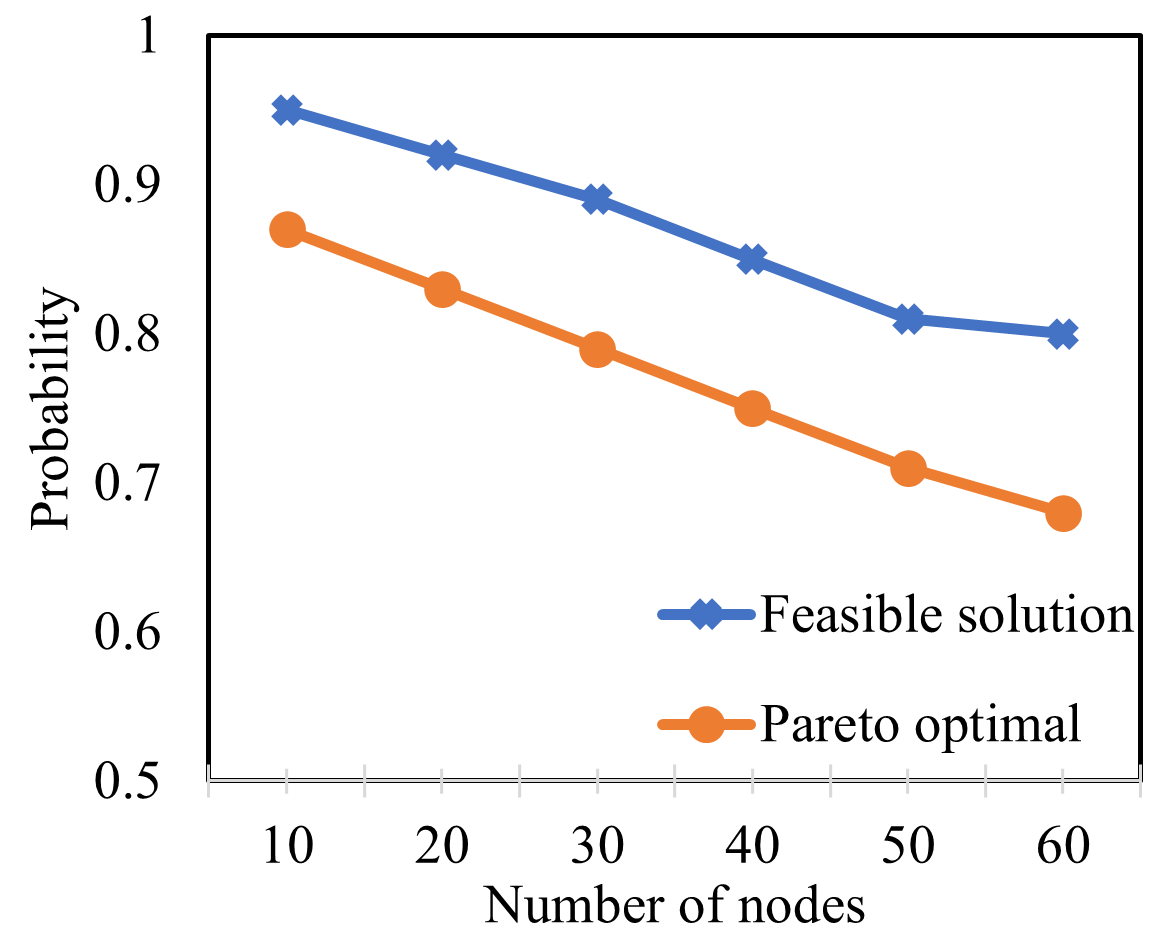}
		\caption{\hspace{0.25cm}BER and hops}
		\label{fig3.3}
	\end{subfigure}
	
	\caption{Probability of feasible solutions and optimal solutions for two objectives.}
	\label{fig3}
\end{figure*}

\section{Experiment and Analysis}
To verify the feasibility of CIM in multi-objective routing problems, we conducted experiments on directed graphs of various sizes, as listed in Table \ref{tab1}. For each size, we randomly generated 40 samples and ran each sample 50 times using the CIM simulator. Some parameters used in the experiments are provided in Table \ref{tab2}.

\begin{table}[htbp]
\caption{Scale of the Experiment}
\begin{center}

	\renewcommand{\arraystretch}{1.5} 
	\resizebox{\columnwidth}{!}{%
		\begin{tabular}{|c|c|c|c|c|c|c|}
			\hline
			Number of Nodes & 10 & 20 & 30 & 40 & 50 & 60 \\ \hline
			Average Number of Edges & 30 & 125 & 280 & 420 & 700 & 1000 \\ \hline
		\end{tabular}
	}
	\label{tab1}
\end{center}
\end{table}

\begin{table}[h]
	\centering
	\caption{Experiment Parameters}
	\renewcommand{\arraystretch}{1.5} 
	\begin{tabular}{|c|c|}
		\hline
		\textbf{Parameters} & \textbf{Values} \\ \hline
		Path loss exponent, $\alpha$ & 2.7 \\ \hline
		Transmitting power, $P_T$ & 50 W \\ \hline
		Carrier wavelength, $\lambda_c$ & 1.2 m \\ \hline
		Modulation scheme & QPSK \\ \hline
	\end{tabular}
	\label{tab2}
\end{table}

\subsection{Simulation Experiment}

We evaluate the performance of CIM in solving multi-objective routing optimization problems under different scenarios: single-objective, two-objective, and three-objective settings. The results demonstrate that CIM consistently finds feasible solutions across various network sizes and topologies, even as the number of nodes and complexity increase. For single-objective optimization, the probability of finding an optimal solution remains high, particularly when objectives have uniform characteristics. In multi-objective scenarios, while the probability of obtaining Pareto optimal solutions decreases with increasing network size, CIM still provides high-quality solutions that are close to the Pareto optimal frontier. 

\subsubsection{Single-Objective Optimization}
Single-Objective Optimization: We initially experimented with CIM using a single objective. Fig. \ref{fig2} presents our experimental results for different objectives: (a) path loss, (b) BER, and (c) number of hops. The graphs illustrate the probabilities of finding feasible and optimal solutions as the number of nodes increases. 

The results indicate that CIM maintains a high probability of finding feasible solutions, even as the network size grows, achieving feasible solutions for networks with up to 60 nodes and approximately 1,000 edges. For path loss and BER, the likelihood of finding optimal solutions decreased with more nodes due to increased complexity, but feasible solutions remained above 80$\%$. In the case of hops (Fig. \ref{fig2.3}), the uniform edge weight led to multiple solutions converging to the ground state, making nearly all feasible solutions optimal. This suggests CIM performs well when the objective function is uniform across the network.

\subsubsection{Two-Objective Optimization}
We extended the experiments to two-objective scenarios, systematically varying the weights from 0.1 to 0.9 to explore a range of Pareto optimal solutions. Fig. \ref{fig3} shows the results for three pairs of objectives: (a) path loss and BER, (b) path loss and hops, and (c) BER and hops, with equal weights (0.5 each).

The experimental findings show that CIM is capable of identifying feasible solutions in the majority of cases, with a substantial proportion of these solutions being Pareto optimal. As the number of nodes increases, the probability of finding Pareto optimal solutions gradually decreases, which is a common challenge in multi-objective optimization due to the increasing complexity of finding solutions that balance multiple conflicting objectives. Nonetheless, the non-Pareto optimal solutions obtained were generally close to the Pareto optimal frontier across different objective dimensions, indicating that CIM can still provide high-quality solutions even when strict optimality is not achieved.

\begin{figure}[htbp]
	\centerline{\includegraphics[scale=0.5]{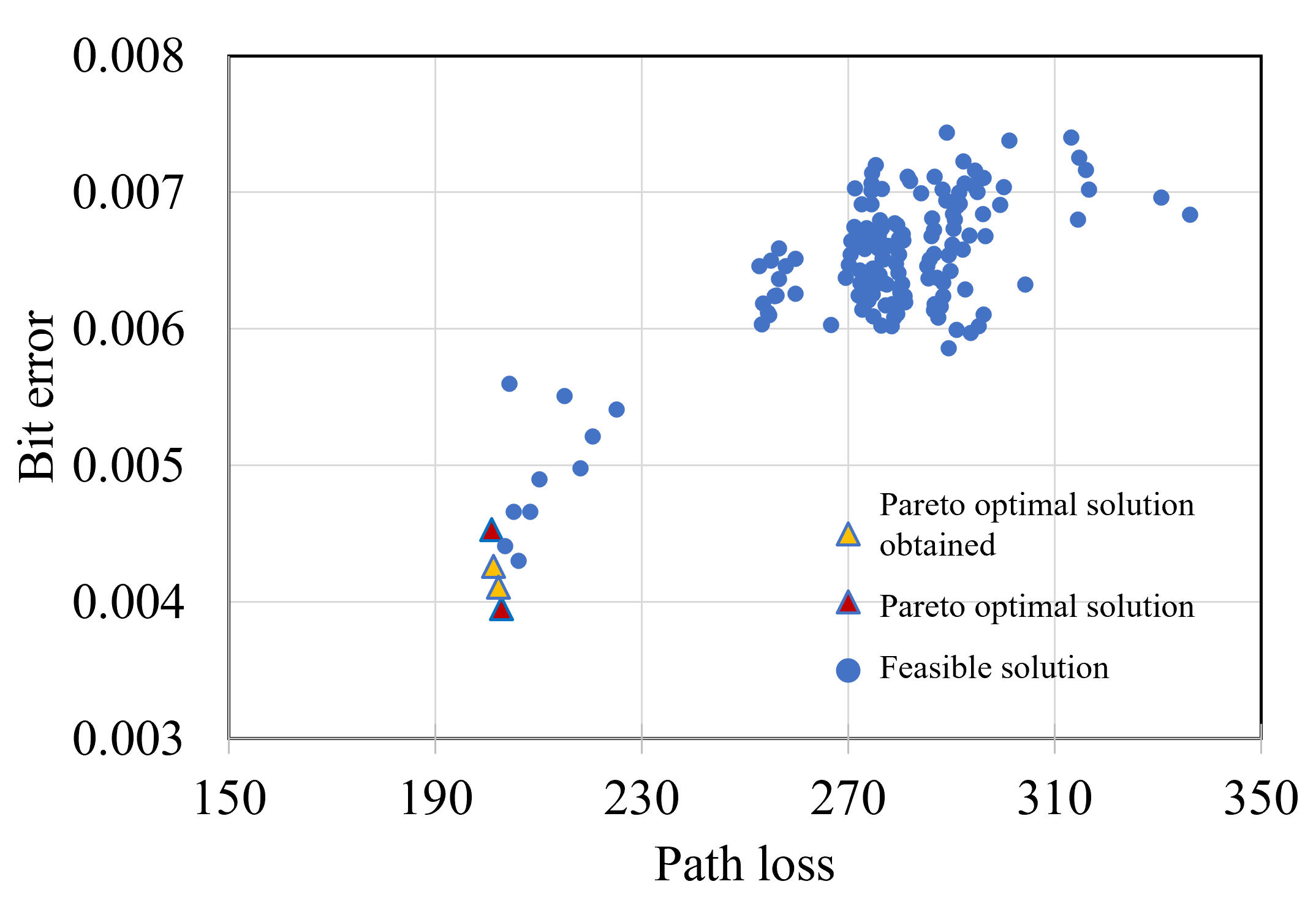}}
	\caption{The solution space of 20 nodes}
	\label{fig4}
\end{figure}

Fig. \ref{fig4} illustrates a detailed view of the solution space for the path loss and BER objectives with equal weighting in a 20-node network. The blue circles represent the top 250 feasible solutions selected from the sample, while the yellow triangles highlight the subset of Pareto-optimal solutions obtained by CIM. The red triangles indicate the full set of Pareto-optimal solutions. As shown, CIM was able to discover a portion of the Pareto-optimal solutions, and these solutions are well-distributed near the Pareto frontier, indicating the algorithm's effectiveness in balancing the trade-offs between path loss and BER.

\subsubsection{Three-Objective Optimization}
\begin{figure}[htbp]
	\centerline{\includegraphics[scale=0.5]{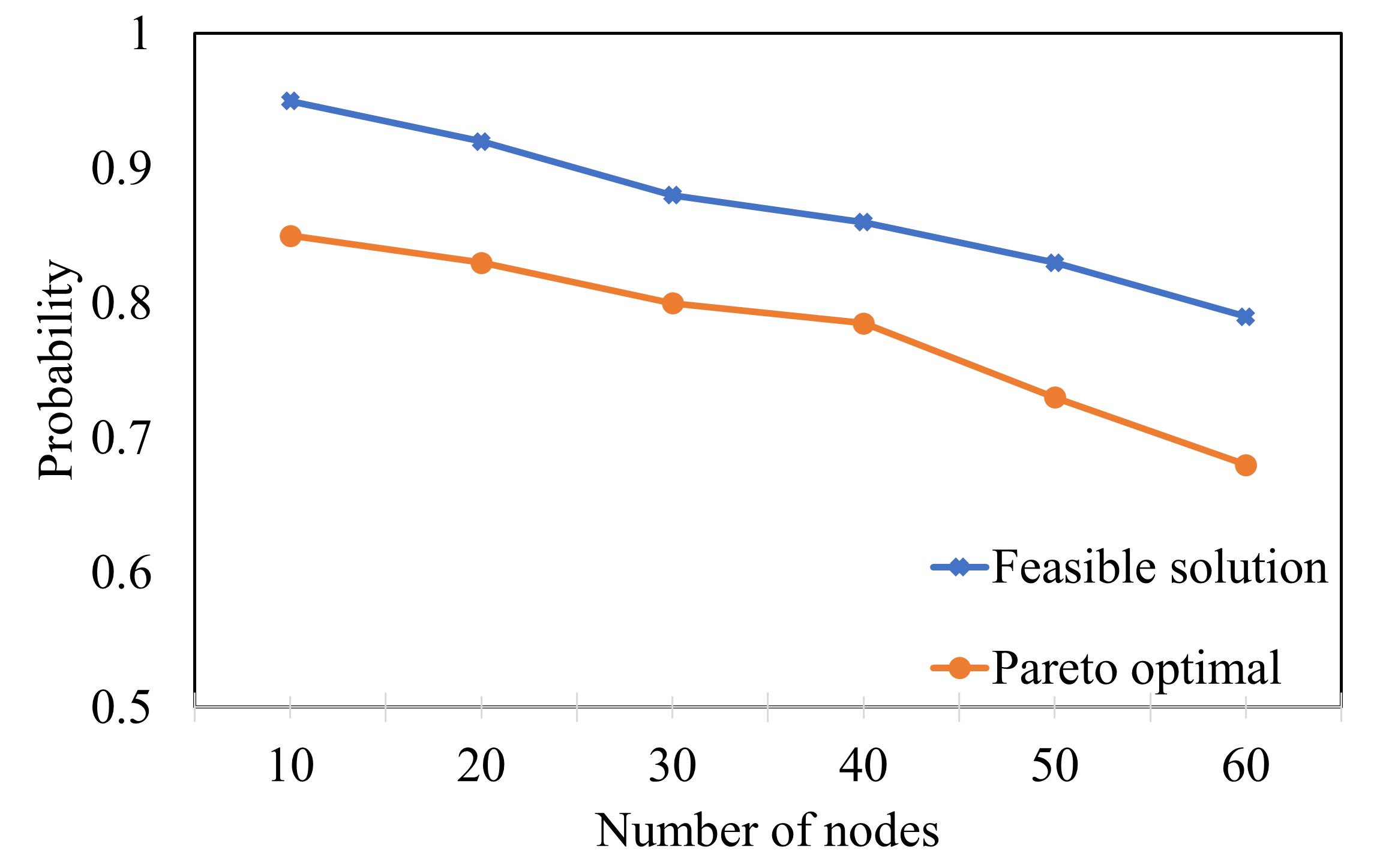}}
	\caption{Probability of feasible solutions and optimal solutions for three objectives.}
	\label{fig5}
\end{figure}

Finally, we conducted experiments with three objectives, varying the weights from 0.1 to 0.8. Fig. \ref{fig5} shows the results when all weights were set equally to 1/3.

The results show that CIM effectively finds feasible solutions across various network sizes. However, as the number of nodes increases, the probability of obtaining Pareto optimal solutions decreases, similar to the trend in the two-objective experiments. Despite the three conflicting objectives, the feasible solutions found by CIM tend to be close to the Pareto optimal set, demonstrating its ability to handle complex multi-objective optimization.

\subsection{Complexity Analysis}
We perform a complexity analysis and comparison of several classical and quantum routing algorithms with CIM, as presented in Table \ref{tab3}. The analysis scenario is set as a weighted positive edge-directed graph\cite{bib13}.

\begin{table}[htbp]
	\caption{Comparison of Algorithm Complexities}
	\begin{center}
		\renewcommand{\arraystretch}{1.5} 
		\setlength{\tabcolsep}{7pt} 
		\begin{tabular}{|>{\centering\arraybackslash}m{3cm}|>{\centering\arraybackslash}m{2.0cm}|>{\centering\arraybackslash}m{1.7cm}|}
			\hline
			\textbf{Algorithm} & \textbf{Time Complexity} & \textbf{Space Complexity} \\
			\hline
			Bellman-Ford & $\mathcal{O}(VE)$ & $\mathcal{O}(V)$ \\
			\hline
			Dijkstra & $\mathcal{O}(V^2)$ & $\mathcal{O}(V)$ \\
			\hline
			Dijkstra with Fibonacci Heap & $\mathcal{O}(V + E \log V)$ & $\mathcal{O}(V)$ \\
			\hline
			Khadiev \& Safina (quantum gate) & $\mathcal{O}(\sqrt{VE} \log V)$ & $\mathcal{O}(V)$ \\
			\hline
			Heiligman (quantum gate)& $\mathcal{O}(V^{7/4})$ & $\mathcal{O}(V)$ \\
			\hline
			CIM-Simulator & $\mathcal{O}\left(E\right)$ & $\mathcal{O}(E^2)$ \\
			\hline
		\end{tabular}
		\label{tab3}
	\end{center}
\end{table}
Table \ref{tab3} presents a complexity comparison between CIM and other algorithms for routing problems. For the CIM-Simulator, the number of cycles of calculating (\ref{label_20}) and (\ref{label_21}) is usually a constant level, so the time complexity of the CIM-Simulator is $O(E)$. This indicates that CIM performs exceptionally well in sparse graphs or networks with fewer connections.

The other quantum algorithms in the table may have varying performances based on specific network structures (such as node density and topology complexity). In contrast, CIM can efficiently work across different network topologies without requiring problem-specific adjustments, showcasing greater adaptability and generality.
This adaptability makes CIM well-suited for large-scale, multi-objective routing optimization problems, and it can provide near-optimal solutions in various scenarios.
\section{Conclusion}
This paper has demonstrated the application of CIM to the task of multi-objective optimization in wireless multi-hop network routing. By reformulating the network routing problem as an Ising model and utilizing the QUBO framework, we have shown how CIM can efficiently manage the intricate trade-offs and constraints present in such systems. Our analysis highlights CIM's potential to address multi-objective optimization problems effectively. The findings suggest that CIM is a promising scalable alternative for other complex optimization challenges in communication networks. Future research could build on these results by refining the model, applying it to various network configurations, and exploring real-world implementations of CIM in other optimization domains.

\section*{Acknowledgment}
The authors gratefully acknowledge the support from the National Natural Science Foundation of China through Grants Nos. 62461160263 and 62371050.

\bibliographystyle{IEEEtran}  
\bibliography{bibgraphy}  

\end{document}